\documentclass[sigconf,natbib=false, nonacm]{acmart}
\AtBeginDocument{%
  }

\setcopyright{acmlicensed}
\copyrightyear{2018}
\acmYear{2018}
\acmDOI{XXXXXXX.XXXXXXX}
\acmConference[ICSE SEIP]{The 48th International Conference on Software Engineering -- Software Engineering in Practice}{April 12--18, 2025}{Rio De Janieiro, BR}
\acmISBN{978-1-4503-XXXX-X/2018/06}

\usepackage{listings}
\usepackage[dvipsnames]{xcolor}
\usepackage{array}
\usepackage{multirow}

\lstdefinelanguage{RAPID}{
  morekeywords={PROC,ENDPROC,MoveJ, MoveL, Stop, Offs},
  sensitive=true,
  morecomment=[l]{!}
}

\lstdefinestyle{RAPIDStyle}{
  language=RAPID,
  float=t,
  floatplacement=t,
  basicstyle=\small\ttfamily,
  moredelim=**[is][\color{red}]{@r}{@},
  moredelim=**[is][\color{green!60!black}]{@g}{@},
  moredelim=**[is][\bfseries]{@f}{@},
}

\lstdefinestyle{RAPIDStyleTable}{
  language=RAPID,
  basicstyle=\small\ttfamily,
  moredelim=**[is][\color{red}]{@r}{@},
  moredelim=**[is][\color{green!60!black}]{@g}{@},
  moredelim=**[is][\bfseries]{@f}{@},
}

\lstdefinestyle{PROMPT}{
  language=Python,
  floatplacement=tpb,
  basicstyle=\small\ttfamily,
  keywordstyle=\color{Black},
  moredelim=**[is][\color{red}]{@r}{@},
  moredelim=**[is][\color{green!60!black}]{@g}{@},
  moredelim=**[is][\bfseries]{@f}{@},
}

\RequirePackage[
  datamodel=acmdatamodel,
  style=acmnumeric,
  ]{biblatex}

\begin{document}

\title{Utilizing LLMs for Industrial Process Automation: A Case Study on Modifying RAPID
Programs}

\author{Salim Fares}
\orcid{0009-0003-3138-3518}
\affiliation{%
  \institution{University of Passau}
  \city{Passau}
  \country{Germany}}
\email{salim.fares@uni-passau.de}

\author{Steffen Herbold}
\orcid{0000-0001-9765-2803}
\affiliation{%
  \institution{University of Passau}
  \city{Passau}
  \country{Germany}
}
\email{steffen.herbold@uni-passau.de}

\renewcommand{\shortauthors}{Fares et al.}

\begin{abstract}
\label{sec:abstract}
How to best use Large Language Models (LLMs) for software engineering is covered in many publications in recent years. However, most of this work focuses on widely-used general purpose programming languages. The utility of LLMs for software within the industrial process automation domain, with highly-specialized languages that are typically only used in proprietary contexts, is still underexplored. Within this paper, we study enterprises can achieve on their own without investing large amounts of effort into the training of models specific to the domain-specific languages that are used. We show that few-shot prompting approaches are sufficient to solve simple problems in a language that is otherwise not well-supported by an LLM and that is possible on-premise, thereby ensuring the protection of sensitive company data. 
\end{abstract}



\maketitle

\section{Introduction}
\label{sec:Introduction}

How to best use Large Language Models (LLMs) has been the hottest topic in software engineering research in the last years, with publications covering a large number of software engineering tasks~\cite{10.1145/3695988} or the efficiency gains (or lack thereof) when developers use LLMs to automatically generate code~\cite{mohamed2025impact}. Usually, this research considers the software development with widely-used general purpose programming languages like Python or Java. These are also the languages, which are most often seen during the training of LLMs due to the vast amounts of data available in public repositories, e.g., on GitHub~\cite{li2023starcodersourceyou}. If and how LLMs can support software development in other technologies, that do not have vasts amounts of open-source data, is still mostly unexplored. 
Within this paper, we contribute to the assessment of the utility of LLMs for software development for industrial automation to support the programming of robotic arms. While LLMs have shown success in generating robot control scripts in general-purpose languages like Python~\cite{huang2023instruct2actmappingmultimodalityinstructions}. These successes not withstanding, domain-specific languages like ABB's RAPID Programming Language (RAPID) are the norm in this domain, making the practical value of success in languages like Python questionable with respect to this domain. Moreover, the proprietary nature of the RAPID language paired with the fact that software is developed for specialized, industrial applications that are often carefully guarded business secrets guiding production processes, means that there is only a low amount of publicly available source code, making it unclear if LLMs trained on such data can provide any benefit at all.
One obvious solution to this problem are specialized LLMs, that huge industrial companies like ABB and Siemens develop, i.e., companies that already serve as defacto providers of standardized domain-specific language and tooling. While such technologies are already emerging~\cite{siemens_industrial_copilot}, the industrial giants still have, to some degree, similar problems with sourcing training data, since the companies that build robotics based on their hard- and software still do not share the source code, making the success of such LLMs uncertain. 

Therefore, we explore a different direction and try to understand if a Small and Medium-sized Enterprise (SME) on its own can already harness existing general-purpose LLMs to assist their development without expensive solutions like training their own models. To this end, we explore the hypothesis that a current mid-sized LLM (in our case Llama 3.1 70B~\cite{meta2024llama31}) can be used with carefully engineering few-shot prompts to support specific, simple but often reoccurring, programming tasks.
We focus on three common programming tasks that aim at modifying existing routines defined in RAPID to program ABB robotic arms. : (1) modifying arguments within existing routines, (2) adding offset instructions to introduce positional shifts, and (3) reversing movement routines. These tasks represent different complexity from simple parameter changes to structural rewrites, and collectively reflect key challenges in robot programming. Moreover, the tasks are practically relevant and frequently re-occurring, e.g., when an existing production line design is modified to fit new requirements. Our research helps us to gain insights into the following two research questions. 

\begin{itemize}
    \item \textbf{RQ1}: Can a general-purpose LLM generate valid RAPID code using only prompt engineering for clearly specified source code modification tasks?
    
    \item \textbf{RQ2}: Does the LLM adhere to all formatting rules and programming guidance when generating RAPID code, when these guidelines are provided in the prompt?

\end{itemize}

We study these questions in collaboration with an SME to not only understand the general feasibility, but rather to assess these questions directly on existing code within the strict coding guidelines of the company. Our results indicate that the model achieves near-perfect accuracy $\approx99\%$ (according to our custom validator and dataset) on low-complexity tasks such as argument modification, which primarily involve pattern recognition and simple renaming. For moderately complex tasks, like adding offsets that require structural changes and insertion of new instructions, the model performs with high accuracy $\approx91\%$, provided it is given multiple attempts to generate the correct output. However, performance declines on high-complexity tasks, such as reversing movement routines that demand rewriting entire logic flows and handling special cases, with accuracy ranging from $\approx77\%$ to $\approx83\%$.

These findings contribute to the field in several key ways.
\begin{itemize}
    \item We demonstrate the practical feasibility of employing general-purpose LLMs to conduct simple modifications of RAPID code through prompt engineering alone, eliminating the need for costly task-specific retraining.
    \item By systematically evaluating performance across tasks of varying complexity, we found that once structural modifications are required (e.g., adding instructions or modifying the order of statements), the error-rate of the general-purpose LLM increases, though the overall accuracy is still relatively high. 
    \item We assess how a non-LLM solution to similar problems could look like and compare this to the LLM solution, showing that both approaches have advantages and drawbacks. 
\end{itemize}   
Overall, our work shows that given a fixed use case with clearly defined boundaries and sufficiently low complexity, the adoption of LLMs to support code generation can be achieved with a relatively low investment, even in niche-domains for languages that are not well-known by the LLM itself.

The remainder of this paper is structured as follows: Section~\ref{sec:related_work} reviews related work on the use of LLMs in robotics and code generation. Section~\ref{sec:data} introduces the use case development and the data used for this study. Section~\ref{sec:methodology} gives an overview of our approach, outlines the prompt-based methodology and output validation. Section~\ref{sec:experiments} describes the experimental setup, while Section~\ref{sec:results} presents the results. Section~\ref{sec:discussion} discusses the findings and answer the research questions. Section~\ref{sec:threats} discusses threats to the validity of our results and Section~\ref{sec:conclusion} concludes the paper.

\section{Related Work}
\label{sec:related_work}
Recent research has increasingly explored the role of AI and LLMs in robotic programming. Fakih \textit{et al.}~\cite{Fakih_2024} introduces LLM4PLC, a user-guided iterative pipeline that uses LLMs to generate Programmable Logic Controller (PLC) code from natural language specifications. Their approach incorporates user feedback and external verification tools to ensure the reliability of the generated code. They further utilize Low-Rank Adaptations (LoRAs) fine-tuning and Prompt Engineering to enhance the generation. Their test results show significant improvement in the $pass@k$ metric (i.e., code generation success rate-as defined in \cite{chen2021evaluatinglargelanguagemodels}.
Furthermore, they used a diverse panel of experts in PLC programming to evaluate the generated code based on a scale from 1 to 10 and predefined criteria: correctness, maintainability, and conformance to industry coding standards (e.g. best practice). 

Antero \textit{et al.}~\cite{robotics13090137} proposed a method to reduce the cost of programming complex robot behaviors by LLMs. Their approach involved a Generator LLM that produced task plans using predefined, human-authored software blocks for Finite State Machines (FSMs). These plans were then reviewed by a Supervisory LLM, which validated the output and suggested corrections if errors were detected. This generator–supervisor loop continued until the plan was error-free or a maximum number of iterations was reached, after which the Generator converted the plan into a JSON representation. The Generator received contextual information about the environment, surrounding objects, their properties, and permitted actions. Based on this input, it attempted to reach a user-defined final state using the available resources. The method was evaluated on 11 manually designed tasks involving multiple actions, successfully generating valid plans for 8 of them.

Morano-Okuno \textit{et al.}~\cite{10823586} present a framework that combines digital twins, AI reasoning, and human–robot interaction models to simplify collaborative robot (cobot) programming for non-experts. Their approach enables users to interactively train cobots within a simulated environment: the AI perceives the virtual workspace, suggests and refines task plans of pre-defined actions, incorporates user feedback, and iteratively tests and adapts behaviors, all without generating executable code. They validated their system using a set of scenario-based simulations; however, the paper does not report detailed quantitative task-specific performance metrics (such as success rates or time‑to‑completion percentages). Instead, the evaluation emphasizes the reduced learning curve and usability benefits of interactive simulation, not the direct synthesis or evaluation of deployable code scripts.

Other contemporary efforts include the use of LLMs like GPT-4 or PaLM to produce robotic control logic, either through fine-tuning on datasets of robot trajectories~\cite{liang2023codepolicieslanguagemodel} or through vision-language-action models~\cite{brohan2023rt2visionlanguageactionmodelstransfer}. These approaches rely on extensive retraining or simulation-based learning environments. Works like Instruct2Act~\cite{huang2023instruct2actmappingmultimodalityinstructions} has demonstrated the potential of LLMs to bridge natural language and robot control by generating executable Python code. In their approach, the LLM synthesizes perception-driven policies that orchestrate vision models (e.g., Segment Anything Model (SAM) and Contrastive Language-Image Pre-training (CLIP)) and control logic to complete tabletop manipulation tasks. These include object picking, rotation, and rearrangement in constrained environments. The authors evaluated their system on 30 tasks from the VIMABench \cite{jiang2023vimageneralrobotmanipulation} benchmark, achieving a success rate of $70\%$ in simulation and $60\%$ in real-world executions, demonstrating the feasibility of using LLM-generated policies for physical robot control. However, the generated scripts are written in Python, a flexible and general-purpose language with access to high-level APIs and perception models. Whereas, industrial robots such as ABB’s rely on proprietary domain-specific languages like RAPID, which impose stricter syntax, semantics, and execution constraints. Instruct2Act does not address this layer of control or the challenges involved in generating such specialized code.

In contrast to prior work focused on simulation-based training or domain-specific tools, our approach investigates whether general-purpose large language models can support code-level interaction with industrial robot languages. Specifically, we explore the capabilities of Llama 3.1 70B to generate and modify RAPID code using prompt engineering alone without fine-tuning or retraining. Given its exposure to a broad corpus of natural language and programming languages during pretraining, we hypothesize that Llama 3.1 may exhibit some generalization to domain-specific languages like RAPID, even if it was not explicitly trained on them. This investigation sets the foundation for evaluating the feasibility of using off-the-shelf LLMs as lightweight tools for industrial robot programming.

\section{Use Case Description and Data}
\label{sec:data}

\subsection{Use case development}

Our work is conducted directly within an an industrial context. We collaborated with AKE Technologies (AKE)\footnote{\url{https://ake-technologies.de/}} a company specialized in developing and manufacturing production systems in the fields of assembly and testing technology, vehicle interior products, noise acoustics and environmental technology. The type of software development conducted at AKE is typical for manufacturing companies that automate their processes and provides a perfect environment to assess how LLMs can aid in this field. 

We developed the use case we study directly together with AKE in an interactive manner: we provided the expertise in collecting data about their software development and the use of LLMs to aide software engineering tasks. Moreover, the first-author of this paper was onboarded into AKE's software development processes. Through this, we established a starting point by understanding which data is available and which programming tasks are suitable targets for our study. 

We identified RAPID programming as suitable technology to be targeted. We identified a need to frequently adopt existing programs, as robotic components are frequently reused across different production lines that are developed. In such cases, developers can often use code from past projects as skeleton for the new robot behavior. In such cases, the old codes needs to be modified to work with the variable names of the new project, as well as often support other movements of the robotic arms. Consequently, we decided to target programming tasks required for the adoption of existing movement routines defined in RAPID.

\subsection{Data}

The main data source are backup files of finished projects from AKE. Each project backup contains the source code for multiple tasks, where each task corresponds to a distinct robotic arm and is implemented through several modules organized by functionality (e.g., variables, functions, and movement routines). This modular design enhances script readability, maintainability, and re-usability across projects. For the purposes of this study, we limited our analysis to modules containing movement routines, as discussed above. A somewhat unusual property of the source code is that it mixes languages: while the keywords of the programming language, as well as any identifier that is defined by the programmers is based on English, the comments to document the code are written in German. 

We collected a total of $466$ backup files from $75$ unique projects. Multiple backups of the same project varied only in positional configurations, while all other elements remained consistent.

From the backups we collected, we extracted \( \mathrm{13776} \) unique movement routines, which we further subdivided into simple movement routines and complex movement routines as shown in Table \ref{tab:data_statistics}.

Simple movement routines mostly involve two positions (start and end) and they share a standard structure (See Listing \ref{lst:simple_movement_routine}), whereas complex movement routines are across more than two positions to handle a special case depending on the project in development (See Listing \ref{lst:complex_movement_routine}):
\begin{figure}[!t]
    \centering
    \begin{lstlisting}[language=RAPID, caption={An example of a simple movement routine.},captionpos=b, label=lst:simple_movement_routine,]
    PROC mvid1_id2()
        !From:  Start Position
        !To:    End Position
        MoveJ id1,position1,velocity,zone,tool\WObj:=world_object\NoMove;
        MoveJ id2,position2,velocity,zone,tool\WObj:=world_object;            
    ENDPROC     
    \end{lstlisting}    
\end{figure}
\begin{figure}[!t]
    \centering
\begin{lstlisting}[language=RAPID, caption={An example of a complex movement routine.}, captionpos=b, label = lst:complex_movement_routine]
PROC special_case_routine()
    MoveJ Offs(position1,0,0,100), velocity, zone, tool;
    MoveL position1, velocity, zone, tool;
    Stop;
    MoveL Offs(position1,0,0,50), velocity, zone, tool;
    MoveJ Offs(position2,0,0,50), velocity, zone, tool;
    MoveL position2, velocity, zone, tool;
    Stop;
    MoveL Offs(position2,0,0,50), velocity, zone, tool;
    MoveJ Offs(position3,0,0,50), velocity, zone, tool;
    MoveL position3, velocity, zone, tool;
    Stop;
    MoveL Offs(position3,0,0,50), velocity, zone, tool;
    MoveJ Offs(position4,0,0,50), velocity, zone, tool;
    MoveL position4, velocity, zone, tool;
ENDPROC
\end{lstlisting}
\end{figure}

Since AKE Technologies standardizes their scripts, all simple routines follow a consistent definition pattern. We used regular expressions to extract the simple routines from each collected file. We then parsed those routines with the help of a reference manual for ABB RAPID Robotics\cite{abb2025rapid} to identify each argument in the instruction. The extracted details were used for both creating prompts and validating the output.

From the extracted unique movement routines, we had \( \mathrm{12196} \) simple movement routines and \( \mathrm{3293} \) complex movement routines as shown in Table \ref{tab:data_statistics}.

\renewcommand{\arraystretch}{1.5}
\begin{table}[!t]
\centering
\caption{Sampled Data}
\label{tab:data_statistics}
\begin{tabular}
{|c|c|}
\hline
Projects & $75$
\\
\hline

Backups& $466$
\\
\hline

Procedures & $18995$
\\
\hline

Simple movement routines & $12196$
\\
\hline

Complex movement routines & $3293$
\\
\hline

Other Procedures & $3506$ 
\\
\hline
\end{tabular}
\end{table}

The executed action in a position determines the other arguments (velocity, zone, robotic tool, work objects, ...) in the movement instructions. Based on that we categorized the positions into four types:
\begin{itemize}
    \item Pre-Position: a passing through position.
    \item Work-Position: a stationary position, where we expect the robot to execute a specific task.
    \item End-Position: the position where the robot finishes a task and must stop there.
    \item Off-Position: any position after applying an offset on it.
\end{itemize}

This study specifically focuses on the analysis of simple movement routines between two Pre-Positions, because they are the simplest routines regarding the definition structure, the low number of special cases, and their frequent occurrence in all projects. This focus supports the development of a custom rule-based validator to verify output correctness. Additionally, it offers practical benefits to AKE developers by reducing repetitive manual validation and modification. We leave the other routine types for future research, though we see no indication why our results should not generalize to the other positions, since we already cover different task complexities.

With 1731 examples for the Pre-Positions routines, we used 11 examples for our prompts and 1720 examples for testing, leaving 10465 examples for the other cases. For simplicity we refer to movement routine and movement instruction as just routine and instruction.

\section{Approach}
\label{sec:methodology}
In this study we focus on modifying RAPID programming scripts. This means that the task input will have a RAPID script and that we do not have to generate the code script from scratch. Based on that we propose an AI-based solution, using an LLM in an inference setting with few-shots prompting, without additional fine-tuning. And we accompany it with a customized rule-based validator to validate the generated output.

\subsection{Overview}
We selected the Llama 3.1 70B model, with its large parameter size, which typically translates to strong language modeling and comprehension of code syntax and semantics \cite{meta2024llama31}. We decided against using a proprietary, cloud-based model due to the sensitive nature of data involved. Moreover, a model with 70B parameters can be used for inference on a single A100, i.e., no complex hardware setup is required by an SME who wants to use this locally within a restricted use case and their own inference server. 

While RAPID is not common in pretraining data (e.g., not even part of the 32 programming languages included in The Stack v2~\cite{lozhkov2024starcoder}), it is structural similarity to other languages for which data is publicly available (e.g., Pascal). Therefore, while we do not have knowledge about the exact data used for training Llama 3.1, we believe it is reasonable to assume that the model knows similar languages to RAPID and should, therefore, be able to have a general understanding of the of RAPID context, be able to refactor code, and even generate logic consistent with the style and constraints of the RAPID language using only prompts.

In this study we consider three tasks:

\begin{enumerate}
    \item \textbf{Modifying routine arguments.} Modify an implemented routine by changing a specific argument for all of its instructions (for example changing the velocity or the number of the station, etc.). We consider this task to have the lowest complexity as it is basically pattern recognition and name changing according to the rules and examples provided in the prompt. The future purpose of this task is to autonomously change all required argument when the routine's type is changed (i.e., from a pre-position routine to a work-position routine).
    \item \textbf{Add an Offset instruction to a routine.} Apply an offset on a specific position in an implemented routine by adding an complete instruction containing the offset function on the position argument. This task has a higher complexity than argument modification, because it requires the model to add a new instruction in the correct order and taking its arguments from user prompt and the routine definition. The future purpose of this task is to autonomously add an offset instruction to a specific pre-position (in special cases) and before reaching or leaving a work-position.
    \item \textbf{Reversing a routine.} Generate the reverse routine of an implemented one. In this task the model has to reverse the order of instructions with possible changes in movements type or even adding intermediate instructions adhering to the provided rules in system prompt. In addition, the model has to change the routine's header as well as re-writing comments according to the new source and destination positions. We consider this task to have the highest complexity among the three tasks, because of all the changes that has to be made. In special cases, re-ordering the instruction and swapping start and end positions are not sufficient to reverse a routine. In those cases some instructions may be added, removed or modified to adhere to reverse logic defined by AKE Technologies
\end{enumerate}

By considering tasks of varying complexity, we aim to demonstrate the model's capabilities and address our research questions.

\subsection{Prompts}
For each tasks we defined a system prompt which included rules and programming guidance provided by AKE Technologies with one example for each rule (if needed), selected from the dataset to make the output adhere to their standards and practices when formulating identifiers, instructions and routines. Listing \ref{lst:system_prompt} shows an excerpt of one the prompts we used. Please note that we cannot share the full prompts including examples due to the sensitive nature of the data. The user prompts (input) varied depending on the task:
\begin{enumerate}
    \item \textbf{Modifying routine arguments.} The original routine and the desired modification (See Table \ref{tab:modifying_routine_arguments_example}).
    \item \textbf{Add an Offset instruction to a routine.} The original routine and how to add the instruction. For this study, we only specify start position and end position. (See Table \ref{tab:adding_offset_example}).
    \item \textbf{Reversing a routine.} The source code for one or more simple routines, that should be reversed. (See Table \ref{tab:reversing_example}).
\end{enumerate}

Since the comment mixes English programming language keywords with German comments, we tested both English and German system and user prompts. This is further motivated by prior research that indicates that using English prompts yields superior performance in comparison to other languages~\cite{zhang-etal-2023-dont}. 

\begin{figure}[!t]
\begin{lstlisting}[language=Python, caption={The system prompt for the argument modification task. [EXAMPLES] represent two examples of the task to be solved, i.e., pairs of the expected input and output.}, captionpos=b, label = lst:system_prompt, style=PROMPT]
- You are an expert in robot programming.
- Give only the OUTPUT, no further explanations.
- This is how to formulate a movement instruction:
    Movement_TYPE TARGET_POSITION, VELOCITY, ZONE, TOOL\\WObj:=WORK_OBJECT;
    EXAMPLE:
    MoveJ pR7_400,vR7_rapid,z50,toR7_active\\WObj:=woR7_Base;
- Do not add any additional instructions.
- If the movement type is Machine_Tending, you must add Machine_Tending_ID as follows:
    Movement_TYPE ID,TARGET_POSITION, VELOCITY, ZONE, TOOL\\WObj:=WORK_OBJECT;
    EXAMPLE:
    MT_MoveJ 400, pR7_400,vR7_rapid,z50,toR7_active\\WObj:=woR7_Base;
- The first movement instruction in a movement routine always has rapid velocity, active tool, and base WObj.
[EXAMPLES]
\end{lstlisting}
\end{figure}

\begin{table*}[t]
\centering
\caption{Modifying routine arguments example}
\label{tab:modifying_routine_arguments_example}
\begin{tabular}{|c|c|}
\hline
\multicolumn{2}{|c|}{User prompt: "Use velocity velocity\_2."} \\
\hline
Input & Output \\
\hline
\begin{minipage}[t]{0.48\textwidth}
\scriptsize
\begin{lstlisting}[language=RAPID, style=RAPIDStyleTable]
PROC mvid1_id2()
    !From:  Start Position
    !To:    End Position
    MoveJ id1,position1,@fvelocity_1@,zone,tool\WObj:=world_object\NoMove;
    MoveJ id2,position2,@fvelocity_1@,zone,tool\WObj:=world_object;            
ENDPROC
\end{lstlisting}
\end{minipage}
&
\begin{minipage}[t]{0.48\textwidth}
\scriptsize
\begin{lstlisting}[language=RAPID, style=RAPIDStyleTable]
PROC mvid1_id2()
    !From:  Start Position
    !To:    End Position
    MoveJ id1,position1,@fvelocity_2@,zone,tool\WObj:=world_object\NoMove;
    MoveJ id2,position2,@fvelocity_2@,zone,tool\WObj:=world_object;            
ENDPROC
\end{lstlisting}
\end{minipage}
\\
\hline
\end{tabular}
\end{table*}

\begin{table*}[t]
\centering
\caption{Adding an Offset instruction to a routine example}
\label{tab:adding_offset_example}
\begin{tabular}{|c|c|}
\hline
\multicolumn{2}{|c|}{User prompt: "Add an intermediate movement using Relative Tool after the start position, with 200 on the Y-Axis"} \\

\hline
Input & Output \\
\hline
\begin{minipage}[t]{0.48\textwidth}
\scriptsize
\begin{lstlisting}[language=RAPID, style=RAPIDStyleTable]
PROC mvid1_id2()
    !From:  Start Position
    
    !To:    End Position
    MoveJ id1,position1,velocity,zone,tool\WObj:=world_object\NoMove;
    MoveJ id2,position2,velocity,zone,tool\WObj:=world_object;            
ENDPROC     
\end{lstlisting}
\end{minipage}
&
\begin{minipage}[t]{0.48\textwidth}
\scriptsize
\begin{lstlisting}[language=RAPID, style=RAPIDStyleTable]
PROC mvid1_id2()
    !From:  Start Position
    !To:    End Position
    MoveJ id1,position1,velocity,zone,tool\WObj:=world_object\NoMove;
    @fMoveJ id_intermediate,RelTool(position1,0,100,0=),velocity,zone,tool\WObj:=world_object\NoMove;@
    MoveJ id2,position2,velocity,zone,tool\WObj:=world_object;            
ENDPROC
\end{lstlisting}
\end{minipage}
\\
\hline
\end{tabular}
\end{table*}

\begin{table*}[t]
\centering
\caption{Reversing a routine example}
\label{tab:reversing_example}
\begin{tabular}{|c|c|}
\hline
Input & Output \\
\hline
\begin{minipage}[t]{0.48\textwidth}
\scriptsize
\begin{lstlisting}[language=RAPID, style=RAPIDStyleTable]
PROC mvidf1_idf2()
    !From:  Start Position
    !To:    End Position
    MoveJ id1,position1,velocity,zone,tool\WObj:=world_object\NoMove;
    MoveJ id2,position2,velocity,zone,tool\WObj:=world_object;            
ENDPROC     
\end{lstlisting}
\end{minipage}
&
\begin{minipage}[t]{0.48\textwidth}
\scriptsize
\begin{lstlisting}[language=RAPID, style=RAPIDStyleTable]
PROC mvid2_id1()
    !From:  End Position
    !To:    Start Position
    MoveJ id2,position2,velocity,zone,tool\WObj:=world_object\NoMove;            
    MoveJ id1,position1,velocity,zone,tool\WObj:=world_object;
ENDPROC
\end{lstlisting}
\end{minipage}
\\
\hline
\end{tabular}
\end{table*}

\subsection{Output Validation}
To ensure the consistency of the generated output and its adherence to AKE Technologies' standards, we developed a custom rule-based validator based on coding guidelines, expected structural aspects and the feedback from AKE developers. This was possible due to the strict coding guidance and well-defined use cases. The checks allow us to ensure that only information that is supposed to be changed is modified (i.e., that the LLMs do not modify code unnecessarily), whether information that should be changed is indeed modified, as well as whether general programming rules are followed. Table~\ref{tab:mistakes} give an overview of the mistakes which can be detected by the validator, including a description of each check. 

\begin{table*}[t]
\centering
\caption{Possible mistakes which can be detected by the validator}
\label{tab:mistakes}
\resizebox{\textwidth}{!}{%
\begin{tabular}{|c|c|>{\raggedright\arraybackslash}p{12cm}|}
\hline
\textbf{Task} & \textbf{Mistakes} & \textbf{Description} \\
\hline

\multirow{3}{*}{\textbf{Argument Modification}} 
& \multirow{2}{*}{Wrong \texttt{ARGUMENT$^{\mathrm{1}}$}} 
& After the modification, the in-prompt specified \texttt{ARGUMENT} was wrongly changed or unchanged in one or more instructions. 
\\

\cline{2-3}
& \texttt{KEY$^{\mathrm{2}}$} was changed 
& Something changed in the definition of the routine 
\\

\hline
\hline
\multirow{7}{*}{\textbf{Adding Offset}} 
& No Offset& No offset instruction was added
\\

\cline{2-3}
&Instruction changed 
& One or more arguments in the instruction were changed
\\

\cline{2-3}
& Wrong position 
& The offset was not applied to the specified position
\\

\cline{2-3}
& \multirow{2}{*}{Wrong function} 
& There are two offset functions: \texttt{Offs()} and \texttt{RelTool()}, and the model did not apply the function specified by the user.
\\

\cline{2-3}
&\texttt{KEY$^{\mathrm{2}}$} was changed & Something changed in the definition of the routine
\\

\hline
\hline
\multirow{7}{*}{\textbf{Reversing}} 
& Wrong reverse logic 
& Source and destination were not swapped or the order of instructions was not correctly reversed
\\

\cline{2-3}
& Leaving \textit{HOME} wrongly 
& Routine should have an intermediate instruction when leaving \textit{HOME}
\\

\cline{2-3}
& Returning \textit{HOME} wrongly 
& Unnecessary intermediate instruction before returning \textit{HOME} was added
\\

\cline{2-3}
& \multirow{2}{*}{Wrong movement type}
& The movement type of the instructions was not adjusted in special cases (ex: Cases involving a combination of linear and non-linear movements)
\\

\cline{2-3}
& \multirow{2}{*}{Mismatching types} 
& After reversing the model wrongly changed the movement type of one or more instructions (in non-special cases)
\\

\cline{2-3}
& Wrong ID 
& Wrong ID for the intermediate instructions.
\\

\cline{2-3}
& Wrong position 
& Wrong position for the intermediate instructions.
\\

\hline
\hline
\multirow{7}{*}{\textbf{All Tasks}} 
& \textit{NoMove} instruction 
& The first instruction did not have a \textit{NoMove} parameter
\\

\cline{2-3}
& Invalid \texttt{IDENTIFIER$^{\mathrm{3}}$} 
& If the \texttt{IDENTIFIER}'s formatting does not adhere to AKE's standard
\\

\cline{2-3}
& More instructions
& More than one instruction was added.
\\

\cline{2-3}
& Less instructions
& One or more instructions were removed from input routine.
\\

\cline{2-3}
& More routines 
& The generated routines are more than the input routines.
\\

\cline{2-3}
& Less routines 
& The generated routines are less than the input routines.
\\

\cline{2-3}
& Wrong default values$^{\mathrm{4}}$ 
& After modification the first instruction did not have default values. 
\\

\hline
\multicolumn{3}{l}{$^{\mathrm{1}}$\texttt{ARGUMENT} $\in \lbrace \texttt{station$^{\mathrm{1.1}}$, id, position, velocity, zone, tool, work object} \rbrace$}
\\
\multicolumn{3}{l}{$^{\mathrm{1.1}}$Station is not actually an argument but it is represented in identifiers formatting}\\

\multicolumn{3}{l}{$^{\mathrm{2}}$\texttt{KEY} $\in \lbrace \text{header of the procedure$^{\mathrm{2.1}}$, source (id \& name), destination (id \& name), start position, end position, tool} \rbrace$}
\\

\multicolumn{3}{l}{$^{\mathrm{2.1}}$When we parsed a routine we composed the name of its reverse and added it to the extracted details, which is why whenever a change in the header is detected}
\\ 
\multicolumn{3}{l}{the validator would also detect a change in reverse name}
\\
\multicolumn{3}{l}{$^{\mathrm{3}}$\texttt{IDENTIFIER} $\in \lbrace \texttt{id, position, velocity, zone, tool, work object} \rbrace$}
\\
\multicolumn{3}{l}{$^{\mathrm{4}}$According to the coding guidance of AKE, the first instruction in a routine should have the default values for \texttt{velocity, zone, tool, work object}}
\\
\end{tabular}%
}
\end{table*}

\section{Experiments}
\label{sec:experiments}

We used Ollama\cite{ollama2025modelfile} to run Llama 3.1-70b-instruct-q4\_0 with the default inferencing parameters:
\begin{itemize}
    \item temperature = $0.8$.
    \item top\_p = $0.9$.
    \item top\_k = $40$.
    \item max\_tokens = $8192$.
    \item num\_ctx = $2048$.
\end{itemize}

Due to the randomized nature of the proposed LLM we generated ten outputs for each input and used the validator to check the correctness. We measure the frequency of the success by measuring how many of the ten generated outputs are correct for a given input, defined as follows:

\begin{equation}
    \mathrm{SuccessRate}(i) = \frac{\left| \text{Correct outputs for input } i \right|}{\left| \text{All outputs for input } i \right|} \times 100\%
\end{equation}

\begin{equation}
\mathrm{Frequency}(x) = \left| \left\{ i \in \text{Inputs} \mid \mathrm{SuccessRate}(i) = x \right\} \right|
\end{equation}

Additionally, we compute accuracy as the percentage of input examples for which at least one of the generated outputs is correct:

\begin{equation}
    \label{eq:accuracy}
    \mathrm{Accuracy} = \frac{\left| \left\{ i \in \text{Inputs} \mid \mathrm{SuccessRate}(i) > 0 \right\} \right|}{|\text{Inputs}|} \times 100\%
\end{equation}
\section{Results}
\label{sec:results}

Table \ref{tab1} shows the accuracy for each task as we have defined it in (\ref{eq:accuracy}) showing that the model performed the best using English prompts at the argument modification task with $99.63\%$ accuracy. At adding an offset task the model achieved an accuracy of $91.97\%$, and at the reversing task $83.72\%$ accuracy.
\begin{table}[!t]
\centering
\caption{The Accuracy for each task}
\label{tab1}
\begin{tabular}{|c|c|c|}
\hline

& \multicolumn{2}{c|}{Accuracy}
\\
\hline

Task& German prompts & English prompts
\\
\hline

Arguments Modification&$99.71\%$&$99.36\%$
\\
\hline

Adding an Offset&$91.86\%$&$91.97\%$
\\
\hline

Reversing&$77.27\%$&$83.72\%$
\\
\hline

\end{tabular}
\end{table}

\begin{figure}[!t]
    \centering
    \includegraphics[width=\columnwidth]{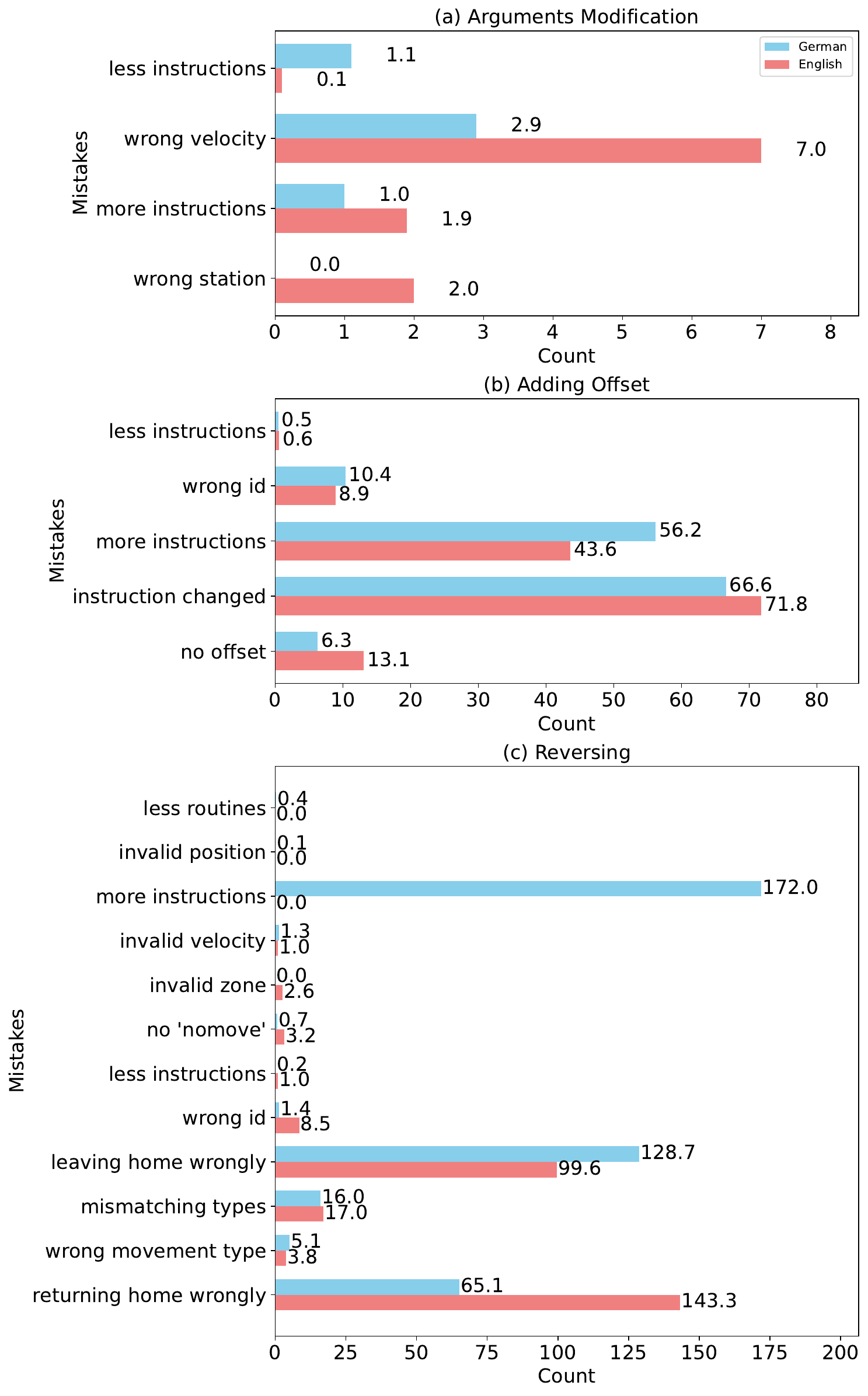}
    \caption{The mistakes detected by the customized validator for the results of each task}
    \label{fig:merged_comparision_mistakes}
\end{figure}

Figure \ref{fig:merged_comparision_mistakes} shows the mistakes detected by the validator for each task for both prompting languages. We only show the mistakes detected for inputs with success rate equal to zero because for the other inputs the model was able to generate at least one correct output. We also divide the occurrences count by ten, since we generate ten outputs for each input and these outputs sometimes have different errors. 
The detected mistakes using English prompts for the argument modification task are only few and manual investigation revealed that they occur in special cases where the name of the position is relatively long, in which case the model changes the other arguments (often the velocity) to match the positions name. As for adding an offset we notice two main mistakes in which the model either modifies some arguments in the original instruction or add unnecessary instructions to the input routine. For the reversing task, we also notice two main mistakes and both of them are related to scenarios involving the HOME position, in which the model has to add or remove an intermediate instruction involving the HOME position according to AKE standards.

\begin{figure}[!t]
        \centering
        \includegraphics[width=\columnwidth]{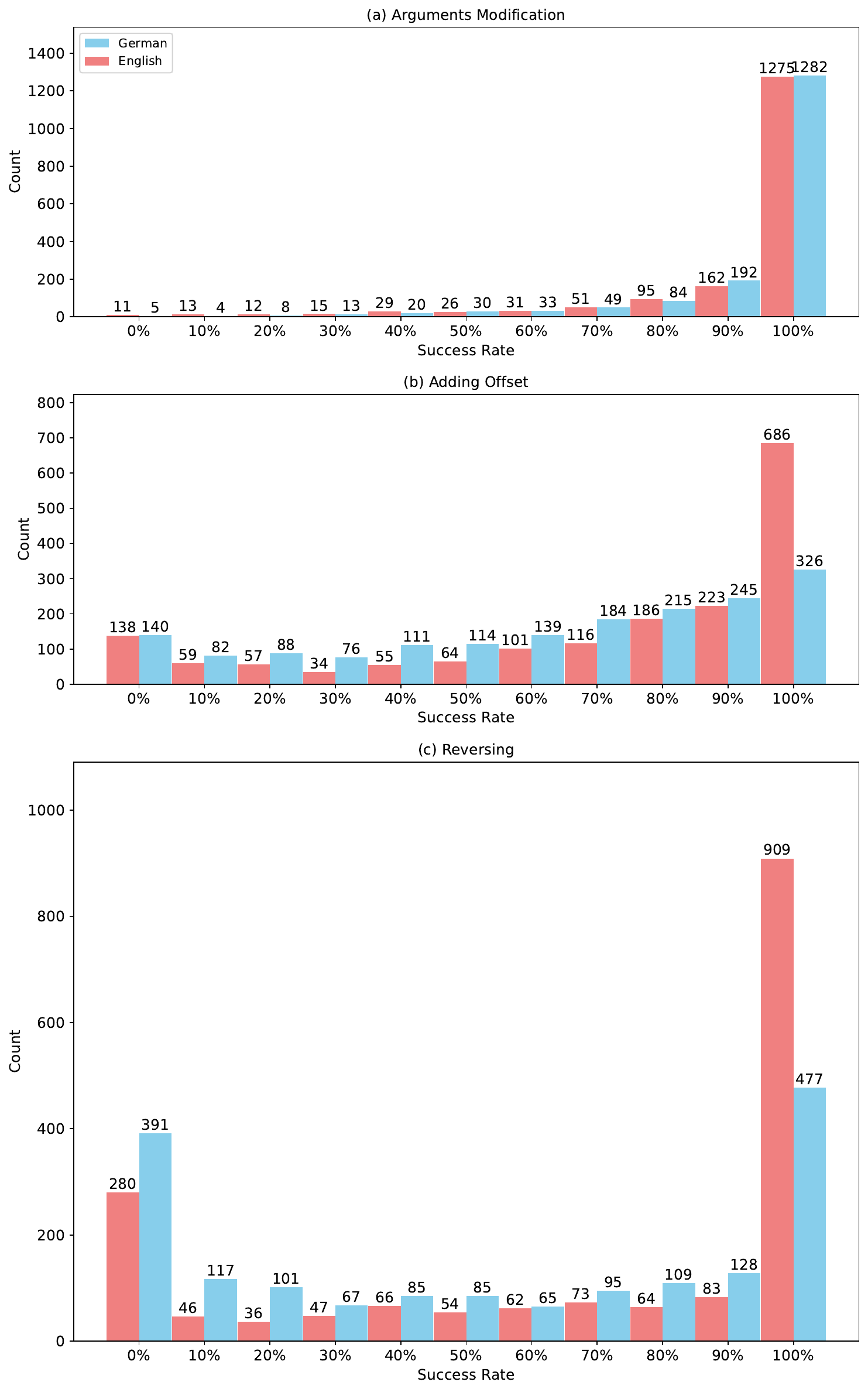}
        \caption{Comparison of success rates in English and German for each task}
        \label{fig:merged_comparision_rates}
\end{figure}

Figure \ref{fig:merged_comparision_rates} shows the count of examples for which the model was able to achieve a specific success rate. We can see that using English prompts the model was able to achieve a success rate of $100\%$ for $\approx74.13\%$ of the input samples at the argument modification task and failed to generate a correct output for $\approx0.64\%$ of the input samples. At adding an offset task, the model achieved a success rate of $100\%$ for $\approx39.88\%$ of the input samples and failed to generate a correct output for $\approx8.03\%$ of the input samples. As for the reversing task, the model achieved a success rate of $100\%$ for $\approx52.85\%$ of the input samples and failed to generate a correct output for around $\approx16.28\%$ of the input samples. These result demonstrate that reliable LLM use for all use cases requires validation rules that enable multiple shots to achieve a correct result.

In comparison, using German prompts achieved similar results for most scenarios. The major difference is noticeable in the reversing task, where the performance dropped from $83.72\%$ using English prompts to $77.27\%$ using German prompts as seen in Table \ref{tab1}. Looking at Figure \ref{fig:merged_comparision_mistakes}, we notice that the model added unnecessary instructions in many cases at the reversing task when using German prompts. Finally the count of input samples, of which the model achieved a $100\%$ success rate, dropped drastically from $\approx39.88\%$ and $\approx52.85\%$ to $\approx18.95\%$ and $\approx27.73\%$ at adding an offset and the reversing task respectively, whereas at argument modification the count was slightly higher. Overall, these results strongly indicate that using English prompts is superior for such code-related tasks.

Figure \ref{fig:merged_comparision_rates} shows the frequency of success rates for each task for both prompting languages.

\section{Discussion}
\label{sec:discussion}

Based on our results, we now answer our research questions focusing on the utility of the LLM-based solution we implemented. Afterwards, we look at the bigger picture and compare our solution to a non-LLM solution for the same tasks and discuss the integration into the daily development process of developers. 

\subsection{RQ1: Can a general-purpose LLM generate valid RAPID code using only prompt engineering for clearly specified source code modification tasks?}

The results support the hypothesis that general-purpose LLMs, even without domain-specific fine-tuning, can perform meaningful and accurate code manipulations in industrial contexts through prompt engineering. Our simple, yet effective, approach is based on few-shot learning such that the LLMs gets a clear pattern of the expected rewrite operations that should be applied to existing code. 

For all three tasks we considered, the few-shot based approach was able to generate valid code that fulfilled the requirements. We find that the complexity of the task is the main driver that affects the LLM performance: while re-writing arguments works almost perfectly, adding instructions at a specific location already fails in about one of ten cases, and re-organizing the complete logic of a function as required for reversing fails in about two of ten cases. 

Based on the structure of the tasks, this suggests that LLMs can reliably handle syntactic-level transformations, particularly when those transformations involve straightforward substitutions, even in niche domains. The more complex tasks reveal that managing logical transformations and sequencing that go beyond token-level substitutions is challenging for the LLM, leading to errors such as adding superfluous instructions or incorrectly applying offsets highlight the importance of precision in industrial code, where even small mistakes can compromise system safety or behavior. This shows that full automation may be possible for the simplest tasks, but anything beyond that requires developers in the loop to check the correctness of the generated code. 

Further, we note that we used a rule-based validation mechanism to catch  mistakes by the LLM and to ensure the quality of the generated code. Our data shows that this not only relevant from a scientific perspective to measure the quality of the generated code, but that such validation mechanisms allow us to greatly increase the accuracy of the LLM-based code generation by discarding invalid results until the LLM is able to generate code that passes the tests. 

\subsection{RQ2: Does the model adhere to all formatting rules and programming guidance, which are provided in the prompt?}

Our detailed error analysis reveals that the models are capable of following detailed instructions how modifications should work, including which rules to follow and what can and cannot be modified. In line with our results for the accuracy of the task completion, LLMs struggle more at following the more detailed instructions of the complex reverse tasks, than of the two simpler tasks. Especially that an intermediary movement instruction is required when leaving home, but unnecessary when going back to the home position, when a movement is reversed led to problems. Moreover, these issues indicate insufficient abstraction or reasoning about procedural flow, which suggests that while the model exhibits strong syntactic fluency, its capacity for higher-level logical inference within domain-constrained code remains limited.

Furthermore, we noted a discrepancy here regarding the language of the prompt in use. The more complex instructions worked a lot better with an English prompt. Hence, even if code is (at least partially) German, describing complex relationships in English -- the language that is typically best trained even in multilingual LLMs~\cite{zhang-etal-2023-dont} -- is beneficial.

\subsection{Comparison to effort required for automation of the tasks without generative AI}

A common best practice before adopting any machine-based solution (including LLMs) is to look for simpler alternatives that do not rely on machine learning~\cite{vertsel2024hybridllmrulebasedapproachesbusiness}. In our case, this could work by, e.g., implementing a rule-based approach that operates directly on the Abstract Syntax Tree, i.e. by using methods known from static analysis of source code. For the renaming of attributes, this is similar to the effort of implementing a rename refactoring, except that additional constraints regarding which names are allowed and how names within a procedure relate to each other need to be fulfilled. Similarly, one could try to define more complex rules that could implement reversing of operations or insertion of movements. Indeed, the definition of such rules is, in principle, comparable to the RegExes we use to validate constraints on the output. 

Given that such an alternative is technically feasible, this raises the question of whether there are possible advantages of LLMs for such a use case or whether the deterministic and engineering static analysis solution should be preferred. From our point of view, the following three aspects could be deciding factors in favor of a LLM: 
\begin{itemize}
    \item \textit{Lack of expertise on the SME's side.} Implementing transformation based on static analysis might be more reliable, but it also requires deep expertise to create suitable ASTs and transform them without breaking programming logic. Depending on whether the developers of the SME have rather a computer science or an engineering background, such expertise might not be available in-house. Engineering suitable prompts, on the other hand, is comparatively easy and rather requires only domain expertise. 
    \item \textit{Simpler corner case handling.} Our results show that the LLM is suitable to solve all problems, including all corner cases like which positions arguments are defined at, without having to specify all of them. While we still need invalidator rules to ensure consistency, defining suitable regular expressions is relatively simple. In comparison, implementing all corner cases and variants for the static analysis requires a dedicated handling of all cases, including possible interactions between combinations, which increases the implementation effort. The static analysis will have problems with any deviation of current standards that might be the case in legacy or future code, whereas our results indicate that the LLM is rather flexible in this regard, demonstrated by the long years of data that could be handled without major issues.     
    \item \textit{Step-stone towards more complex cases.} While the scenarios we cover are sufficiently simple to be possibly engineered with a static analysis approach, more complex settings, in which, e.g., multiple transformations have to be applied in parallel, cannot be solved with static analysis without investing a huge amount of effort anymore. Arguably, LLMs may even be suited to address more complex use cases and the lessons learned from simpler scenarios -- including how to implement them in an organization (see Section~\ref{sec:adoption}) -- can be used to get both acceptance for LLM support by developers, as well as bootstrap the infrastructure and expertise for more complex use cases. 
\end{itemize}

\subsection{Integration into the development process}
\label{sec:adoption}

Regardless of whether done with static analysis or LLMs, the question for both approaches to automatically rewrite code would be how to best integrate them into the development process. In all cases, the developer currently still needs to define herself how to rewrite code, e.g., which parameters need to be renamed, which offsets need to be added, and which movements need to be reversed. 

From our perspective, this is a well-known problem from Integrated Development Environment (IDE) design that is essentially the same as for automated refactorings: while renaming a simple local variable might not make sense with IDE support, renaming a class that is reused across source code automatically is a great benefit. Thus, the efficiency is a matter of the scale of changes that are done automatically without developer interaction. 

Thus, we need to extend the use case from what is changed within the source code to how developers trigger these changes, as confirmed also by AKEs developers who are skeptic regarding efficiency improvements of any simple rewrites - regardless of LLM-based or with static analysis. However, they also outlined how these new features can be the building block for a more powerful use case, guided by LLMs: before the development of the robotic software can start, detailed schedules for the interaction between the machines are created, they are then used by the developers to identify which robots need to move where and when, typically in some mixture of natural language and tabular data. By extending the automated analysis with LLMs from only the source code to include these schedules, we can understand what needs to be rewritten: which kind of movement for which kind of robot, at which timestep. However, whether LLMs are sufficiently reliable to handle this more complex information still needs to be determined in future work. 

\section{Threats to validity}
\label{sec:threats}

The type of experiment we use can be categorized as experimental simulation~\cite{stol2020guidelines}, i.e., we focus on a specific setting in which we have control, favoring the internal validity of our results instead of using a diverse number of settings with less control, which would favor generalizability. Consequently, we have high confidence that our results are valid and generalize to similar settings, i.e., RAPID modification tasks of similar difficulty. However, we cannot directly generalize from our data to domain-specific languages other than RAPID, nor can we generalize to more difficult tasks. Moreover, we conducted our experiments together with an organization that has strict coding guidelines. Our results may not transfer to settings without such guidelines. However, we note that such guidelines are common in robotics, due to the usually safety-critical nature of the involved machines. 

\section{Conclusion}
\label{sec:conclusion}
Within this paper, we have demonstrated that LLMs can be used to solve (simple) programming tasks, even in domains in which they lack training data, simply by providing few-shot prompts that enable the LLM to solve the task through pattern matching. We demonstrated this through three modification tasks of different difficulty for industrial robot programming. Our results show that simple tasks involving identifiers are unproblematic and that even more complex tasks, that require re-writing the logical behavior (in our case, reversing movements) could often be solved by the LLMs. 
These results suggest that LLMs can be valuable assistants in industrial programming workflows for well-defined and structured modifications. However, as our results also show that performance decreases with tasks becoming more complex, we believe that tasks that require deeper program logic understanding or global structural manipulation cannot be solved with such a simple few-shot approach and require further research.

Our current work is limited to assessing the feasibility of LLMs in this industrial context. The still open question is whether using LLMs for this purpose can actually improve the efficiency of the development, or whether non-LLM solutions (e.g., based on the manipulation of the abstract syntax tree) are better suited, or if no automation is necessary at all since human developers already solve the tasks efficiently. This will depend to a large degree on the question of to which degree the automation can be bundled, such that large amounts of code can be modified automatically at the same time: for single short blocks, using the automation might consume more time, while for long or even multiple files at once, there may be speed-ups. We will answer these questions in our future work, where we will integrate the automation capabilities directly into the development process of a company, develop concrete scenarios for their use, and measure the impact on efficiency. 

\begin{acks}
This work was carried out as part of the PLC-GPT project, a collaboration between the University of Passau and AKE Technologies funded by the Bavarian State Ministry of Science and Arts. We are grateful to AKE Technologies to grant us the access to their data and the feedback their engineers regarding the utility that guided the use case development and assessment.
\end{acks}

\printbibliography

@inproceedings{Fakih_2024, series={ICSE-SEIP ’24},
   title={LLM4PLC: Harnessing Large Language Models for Verifiable Programming of PLCs in Industrial Control Systems},
   url={http://dx.doi.org/10.1145/3639477.3639743},
   DOI={10.1145/3639477.3639743},
   booktitle={Proceedings of the 46th International Conference on Software Engineering: Software Engineering in Practice},
   publisher={ACM},
   author={Fakih, Mohamad and Dharmaji, Rahul and Moghaddas, Yasamin and Quiros, Gustavo and Ogundare, Oluwatosin and Al Faruque, Mohammad Abdullah},
   year={2024},
   month=apr, pages={192–203},
   collection={ICSE-SEIP ’24} }

@Article{robotics13090137,
AUTHOR = {Antero, Unai and Blanco, Francisco and Oñativia, Jon and Sallé, Damien and Sierra, Basilio},
TITLE = {Harnessing the Power of Large Language Models for Automated Code Generation and Verification},
JOURNAL = {Robotics},
VOLUME = {13},
YEAR = {2024},
NUMBER = {9},
ARTICLE-NUMBER = {137},
URL = {https://www.mdpi.com/2218-6581/13/9/137},
ISSN = {2218-6581},
DOI = {10.3390/robotics13090137}
}

@manual{abb2025rapid,
  title        = {Technical Reference Manual – RAPID Instructions, Functions and Data Types},
  author       = {{ABB Robotics}},
  organization = {ABB},
  type         = {Technical Reference Manual},
  number       = {3HAC050917-001},
  year         = {2025},
  note         = {RobotWare~6.16},
  url          = {https://search.abb.com/library/Download.aspx?DocumentID=3HAC050917-001},
}

@INPROCEEDINGS{10823586,
  author={Morano-Okuno, Hector Rafael and Sandoval-Benitez, Guillermo and Caltenco-Castillo, Rafael},
  booktitle={2024 International Conference on Electrical, Communication and Computer Engineering (ICECCE)}, 
  title={Using AI and Digital Simulations to Expedite the Learning and Programming of COBOTS}, 
  year={2024},
  volume={},
  number={},
  pages={1-6},
  doi={10.1109/ICECCE63537.2024.10823586}}

@misc{liang2023codepolicieslanguagemodel,
      title={Code as Policies: Language Model Programs for Embodied Control}, 
      author={Jacky Liang and Wenlong Huang and Fei Xia and Peng Xu and Karol Hausman and Brian Ichter and Pete Florence and Andy Zeng},
      year={2023},
      eprint={2209.07753},
      archivePrefix={arXiv},
      primaryClass={cs.RO},
      url={https://arxiv.org/abs/2209.07753}, 
}

@misc{huang2023instruct2actmappingmultimodalityinstructions,
      title={Instruct2Act: Mapping Multi-modality Instructions to Robotic Actions with Large Language Model}, 
      author={Siyuan Huang and Zhengkai Jiang and Hao Dong and Yu Qiao and Peng Gao and Hongsheng Li},
      year={2023},
      eprint={2305.11176},
      archivePrefix={arXiv},
      primaryClass={cs.RO},
      url={https://arxiv.org/abs/2305.11176}, 
}

@misc{brohan2023rt2visionlanguageactionmodelstransfer,
      title={RT-2: Vision-Language-Action Models Transfer Web Knowledge to Robotic Control}, 
      author={Anthony Brohan and Noah Brown and Justice Carbajal and Yevgen Chebotar and Xi Chen and Krzysztof Choromanski and Tianli Ding and Danny Driess and Avinava Dubey and Chelsea Finn and Pete Florence and Chuyuan Fu and Montse Gonzalez Arenas and Keerthana Gopalakrishnan and Kehang Han and Karol Hausman and Alexander Herzog and Jasmine Hsu and Brian Ichter and Alex Irpan and Nikhil Joshi and Ryan Julian and Dmitry Kalashnikov and Yuheng Kuang and Isabel Leal and Lisa Lee and Tsang-Wei Edward Lee and Sergey Levine and Yao Lu and Henryk Michalewski and Igor Mordatch and Karl Pertsch and Kanishka Rao and Krista Reymann and Michael Ryoo and Grecia Salazar and Pannag Sanketi and Pierre Sermanet and Jaspiar Singh and Anikait Singh and Radu Soricut and Huong Tran and Vincent Vanhoucke and Quan Vuong and Ayzaan Wahid and Stefan Welker and Paul Wohlhart and Jialin Wu and Fei Xia and Ted Xiao and Peng Xu and Sichun Xu and Tianhe Yu and Brianna Zitkovich},
      year={2023},
      eprint={2307.15818},
      archivePrefix={arXiv},
      primaryClass={cs.RO},
      url={https://arxiv.org/abs/2307.15818},
}

@online{ollama2025modelfile,
  author       = {Ollama},
  title        = {Valid Parameters and Values for Model Inference},
  year         = {2025},
  url          = {https://github.com/ollama/ollama/blob/main/docs/modelfile.md#valid-parameters-and-values},
  note         = {Accessed July 14, 2025}
}

@misc{chen2021evaluatinglargelanguagemodels,
      title={Evaluating Large Language Models Trained on Code}, 
      author={Mark Chen and Jerry Tworek and Heewoo Jun and Qiming Yuan and Henrique Ponde de Oliveira Pinto and Jared Kaplan and Harri Edwards and Yuri Burda and Nicholas Joseph and Greg Brockman and Alex Ray and Raul Puri and Gretchen Krueger and Michael Petrov and Heidy Khlaaf and Girish Sastry and Pamela Mishkin and Brooke Chan and Scott Gray and Nick Ryder and Mikhail Pavlov and Alethea Power and Lukasz Kaiser and Mohammad Bavarian and Clemens Winter and Philippe Tillet and Felipe Petroski Such and Dave Cummings and Matthias Plappert and Fotios Chantzis and Elizabeth Barnes and Ariel Herbert-Voss and William Hebgen Guss and Alex Nichol and Alex Paino and Nikolas Tezak and Jie Tang and Igor Babuschkin and Suchir Balaji and Shantanu Jain and William Saunders and Christopher Hesse and Andrew N. Carr and Jan Leike and Josh Achiam and Vedant Misra and Evan Morikawa and Alec Radford and Matthew Knight and Miles Brundage and Mira Murati and Katie Mayer and Peter Welinder and Bob McGrew and Dario Amodei and Sam McCandlish and Ilya Sutskever and Wojciech Zaremba},
      year={2021},
      eprint={2107.03374},
      archivePrefix={arXiv},
      primaryClass={cs.LG},
      url={https://arxiv.org/abs/2107.03374}, 
}

@misc{meta2024llama31,
  author       = {Meta AI},
  title        = {{Llama 3.1}: Open-Weight Foundation Models},
  url = {https://llama.com/models/llama-3/},
  year         = {2024},
  note         = {Accessed: Jul. 31, 2025}
}

@misc{jiang2023vimageneralrobotmanipulation,
      title={VIMA: General Robot Manipulation with Multimodal Prompts}, 
      author={Yunfan Jiang and Agrim Gupta and Zichen Zhang and Guanzhi Wang and Yongqiang Dou and Yanjun Chen and Li Fei-Fei and Anima Anandkumar and Yuke Zhu and Linxi Fan},
      year={2023},
      eprint={2210.03094},
      archivePrefix={arXiv},
      primaryClass={cs.RO},
      url={https://arxiv.org/abs/2210.03094}, 
}

@article{10.1145/3695988,
author = {Hou, Xinyi and Zhao, Yanjie and Liu, Yue and Yang, Zhou and Wang, Kailong and Li, Li and Luo, Xiapu and Lo, David and Grundy, John and Wang, Haoyu},
title = {Large Language Models for Software Engineering: A Systematic Literature Review},
year = {2024},
issue_date = {November 2024},
publisher = {Association for Computing Machinery},
address = {New York, NY, USA},
volume = {33},
number = {8},
issn = {1049-331X},
url = {https://doi.org/10.1145/3695988},
doi = {10.1145/3695988},
journal = {ACM Trans. Softw. Eng. Methodol.},
month = dec,
articleno = {220},
numpages = {79}
}

@article{mohamed2025impact,
  title={The Impact of LLM-Assistants on Software Developer Productivity: A Systematic Literature Review},
  author={Mohamed, Amr and Assi, Maram and Guizani, Mariam},
  journal={arXiv preprint arXiv:2507.03156},
  year={2025}
}

@misc{li2023starcodersourceyou,
      title={StarCoder: may the source be with you!}, 
      author={Raymond Li and Loubna Ben Allal and Yangtian Zi and Niklas Muennighoff and Denis Kocetkov and Chenghao Mou and Marc Marone and Christopher Akiki and Jia Li and Jenny Chim and Qian Liu and Evgenii Zheltonozhskii and Terry Yue Zhuo and Thomas Wang and Olivier Dehaene and Mishig Davaadorj and Joel Lamy-Poirier and João Monteiro and Oleh Shliazhko and Nicolas Gontier and Nicholas Meade and Armel Zebaze and Ming-Ho Yee and Logesh Kumar Umapathi and Jian Zhu and Benjamin Lipkin and Muhtasham Oblokulov and Zhiruo Wang and Rudra Murthy and Jason Stillerman and Siva Sankalp Patel and Dmitry Abulkhanov and Marco Zocca and Manan Dey and Zhihan Zhang and Nour Fahmy and Urvashi Bhattacharyya and Wenhao Yu and Swayam Singh and Sasha Luccioni and Paulo Villegas and Maxim Kunakov and Fedor Zhdanov and Manuel Romero and Tony Lee and Nadav Timor and Jennifer Ding and Claire Schlesinger and Hailey Schoelkopf and Jan Ebert and Tri Dao and Mayank Mishra and Alex Gu and Jennifer Robinson and Carolyn Jane Anderson and Brendan Dolan-Gavitt and Danish Contractor and Siva Reddy and Daniel Fried and Dzmitry Bahdanau and Yacine Jernite and Carlos Muñoz Ferrandis and Sean Hughes and Thomas Wolf and Arjun Guha and Leandro von Werra and Harm de Vries},
      year={2023},
      eprint={2305.06161},
      archivePrefix={arXiv},
      primaryClass={cs.CL},
      url={https://arxiv.org/abs/2305.06161}, 
}

@misc{siemens_industrial_copilot,
  author       = {Siemens AG},
  title        = {Siemens Industrial Copilot: Generative AI-Powered Assistant for Industrial Automation},
  year         = {2025},
  url          = {https://www.siemens.com/global/en/products/automation/topic-areas/industrial-ai/industrial-copilot.html},
  note         = {Accessed: 2025-09-22}
}

@inproceedings{zhang-etal-2023-dont,
    title = "Don{'}t Trust {C}hat{GPT} when your Question is not in {E}nglish: A Study of Multilingual Abilities and Types of {LLM}s",
    author = "Zhang, Xiang  and
      Li, Senyu  and
      Hauer, Bradley  and
      Shi, Ning  and
      Kondrak, Grzegorz",
    editor = "Bouamor, Houda  and
      Pino, Juan  and
      Bali, Kalika",
    booktitle = "Proceedings of the 2023 Conference on Empirical Methods in Natural Language Processing",
    month = dec,
    year = "2023",
    address = "Singapore",
    publisher = "Association for Computational Linguistics",
    url = "https://aclanthology.org/2023.emnlp-main.491/",
    doi = "10.18653/v1/2023.emnlp-main.491",
    pages = "7915--7927"
}

@misc{vertsel2024hybridllmrulebasedapproachesbusiness,
      title={Hybrid LLM/Rule-based Approaches to Business Insights Generation from Structured Data}, 
      author={Aliaksei Vertsel and Mikhail Rumiantsau},
      year={2024},
      eprint={2404.15604},
      archivePrefix={arXiv},
      primaryClass={cs.CL},
      url={https://arxiv.org/abs/2404.15604}, 
}

@incollection{stol2020guidelines,
  title={Guidelines for conducting software engineering research},
  author={Stol, Klaas-Jan and Fitzgerald, Brian},
  booktitle={Contemporary Empirical Methods in Software Engineering},
  pages={27--62},
  year={2020},
  publisher={Springer}
}

@article{lozhkov2024starcoder,
  title={Starcoder 2 and the stack v2: The next generation},
  author={Lozhkov, Anton and Li, Raymond and Allal, Loubna Ben and Cassano, Federico and Lamy-Poirier, Joel and Tazi, Nouamane and Tang, Ao and Pykhtar, Dmytro and Liu, Jiawei and Wei, Yuxiang and others},
  journal={arXiv preprint arXiv:2402.19173},
  year={2024}
}
\end{document}